\documentclass[default,iicol]{sn-jnl}

\usepackage{graphicx}%
\usepackage{multirow}%
\usepackage{amsmath,amssymb,amsfonts}%
\usepackage{amsthm}%
\usepackage{mathrsfs}%
\usepackage[title]{appendix}%
\usepackage[dvipsnames]{xcolor}%
\usepackage{textcomp}%
\usepackage{manyfoot}%
\usepackage{booktabs}%
\usepackage{algorithm}%
\usepackage{algorithmicx}%
\usepackage{algpseudocode}%
\usepackage{listings}%
\usepackage{siunitx}
\usepackage{braket}
\usepackage{verbatim}
\usepackage{tikz}
\usepackage{stackengine}
\usepackage{multicol}

\theoremstyle{thmstyleone}%
\theoremstyle{thmstyletwo}%

\theoremstyle{thmstylethree}%

\begin{document}

\title[Article Title]{A cavity-microscope for micrometer-scale control of atom-photon interactions}


\author{\fnm{F.} \sur{Orsi $^{1,\ast}$}}
\author{\fnm{N.} \sur{Sauerwein$^{1}$}}
\author{\fnm{R. P.} \sur{Bhatt$^{1}$}}
\author{\fnm{J.} \sur{Faltinath$^{1, }$}}
\author{\fnm{E.} \sur{Fedotova$^{1}$}}
\author{\fnm{N.} \sur{Reiter$^{1, \dagger }$}}
\author{\fnm{T.} \sur{Cantat-Moltrecht$^{1, \ddagger}$}}
\author{\fnm{J.P.} \sur{Brantut$^{1}$}}

\affil[1]{\orgdiv{Institute of Physics and Center for Quantum Science and Engineering}, \orgname{ Ecole Polytechnique Fédérale de Lausanne (EPFL)}, \city{Lausanne}, \country{Switzerland}}

\abstract{Cavity quantum electrodynamics offers the possibility to observe and control the motion of few or individual atoms, enabling the realization of various quantum technological tasks such as quantum-enhanced metrology or quantum simulation of strongly-correlated matter. A core limitation of these experiments lies in the mode structure of the cavity field, which is hard-coded in the shape and geometry of the mirrors. As a result, most applications of cavity QED trade spatial resolution for enhanced sensitivity. Here, we propose and demonstrate a cavity-microscope device capable of controlling in space and time the coupling between atoms and light in a single-mode high-finesse cavity, reaching a spatial resolution an order-of-magnitude lower than the cavity mode waist. This is achieved through local Floquet engineering of the atomic level structure, imprinting a corresponding atom-field coupling. We illustrate this capability by engineering micrometer-scale coupling, using cavity-assisted atomic measurements and optimization. Our system forms an optical device with a single optical axis and has the same footprint and complexity as a standard Fabry-Perot cavity or confocal lens pair, and can be used for any atomic species. This technique opens a wide range of perspectives from ultra-fast, cavity-enhanced mid-circuit readout to the quantum simulation of fully connected models of quantum matter such as the Sachdev-Ye-Kitaev model.}


\maketitle

\section{Introduction}
\label{sec1}


Cavity quantum electrodynamics (QED) is a central tool of atomic, molecular and optical physics, allowing to realize a variety of quantum engineering tasks. For example, in the strong-coupling regime, atoms can be made to exchange photons with each other via the cavity modes faster than photons get lost \cite{Haroche:2006aa} realizing a synthetic long-range interaction that allows for the quantum simulation of a new class of interacting quantum systems \cite{mivehvar:2021aa, PhysRevX.8.011002,periwal:2021uo, kongkhambut:2022aa, dreon:2022aa, young:2024aa}. It is also the workhorse of quantum measurements and sensing, allowing for quantum-enhanced sensing \cite{Leroux:2010aa, Chen:2011aa, Hosten:2016aa, huang:2023aa, robinson:2024aa}. Already two decades ago, landmark experiments showed for example the real-time tracking of the position of an atom in an `atom-cavity microscope' \cite{Kimble00,pinkse:2000ab, PhysRevLett.98.233601, Khudaverdyan_2008}. These capabilities make cavity quantum electrodynamics platforms a very promising candidate for readout, feedback and networking in future atom-based quantum computing architectures \cite{Kimble:2008aa,Reiserer:2015aa,ramette:2022va, covey:2023aa, li:2024aa}.

\let\thefootnote\relax\footnotetext{*\url{francesca.orsi@epfl.ch}}

An optical cavity operates by setting boundaries for the electromagnetic field, singling out a particular set of modes. The spatial profile of these modes is inherited from the geometry of the mirror assembly. Thus, for all cavity QED experiments the spatial dependence of the atom-cavity interaction is tied to a particular design, determined once and for all at the assembly of the system. Recent progress has allowed post-assembly tuning of the cavity geometry to reach particular mode frequency spacing or even high mode degeneracy in the confocal situation \cite{Kollár_2015,nguyen:2018wr,vaneecloo:2022aa}. However, this tuning is still limited by the spatial structure of the modes, and is accomplished by physically moving the cavity mirrors, at speeds not commensurate with the dynamics of the atoms. Alternatively, physically moving individual trapped atoms in and out of the cavity has allowed to deterministically vary the light-atom couplings \cite{khudaverdyan:2008aa, Reiserer:2013ab, PhysRevLett.128.083201,dordevic:2021aa,deist:2022ab,liu:2023aa}. This is however not adapted to Bose-Einstein condensates or degenerate Fermi gases, where the motion of a single atom is not controlled.


In this work, we introduce a hybrid cavity-microscope device combining the strong light-matter coupling of cavity QED with local addressing using high-resolution optics to reach micrometer-scale and time-resolved control over the coupling between atoms and photons, lifting the constraints set by the mode geometry. This is achieved using a control laser beam that directly modulates locally the light-matter coupling strength. In essence, our system takes the well known capability to project optical structures onto atoms, used for instance in quantum gas microscopes \cite{Weitenberg:2011aa,Zupancic:2016ab}, atomtronics transport experiments \cite{amico:2022aa} or tweezer arrays \cite{PhysRevX.4.021034}, and turns it into the spatial shaping of light-matter coupling. 

In the following, we start with describing the device itself, based on optically contacted high-reflectivity cavity mirrors and aspherical lenses sharing a common optical axis. This geometry guarantees the relative alignment and reduces the additional technical complexity, making it compatible with standard cavity QED or aspherical lens pairs used in existing setups. We then demonstrate the operation of this device on a mesoscopic atomic ensemble of \qty{300} laser-cooled $^6$Li atoms, trapped inside the mode of the cavity, leveraging spectral engineering of the atomic transitions through Floquet dressing. We present the engineering of light-matter coupling through the cavity microscope in two steps, first using the atoms as a high-sensitivity sensor for the control light wavefront, allowing for diagnostic and feedback optimization of the beam shape. Second, we produce a sharp control field capable of addressing the atoms at the micrometer scale, and use it in a scanning probe measurement to reconstruct the density distribution of a gas fully contained in the mode of the cavity, combining known super-resolution cold-atoms methods \cite{Hausler:2017ab,mcdonald:2019aa,subhankar:2019aa,veyron:2023aa} with the unique features of cavity QED, similar to recent proposals \cite{Yang:2018aa,Yang:2018ab}. This method allows to program the light-matter coupling, offering more flexbility for local adressing than the use of multimode cavities, however without the benefit of enhanced cooperativiy \cite{PRXQuantum.4.020326, PhysRevLett.88.043601}. These capabilities will open the possibility to program the cavity-induced photon-exchange interaction in space and in time  \cite{periwal:2021uo,kroeze2023replica,Sauerwein_2023}, a promising platform for analogue quantum simulation \cite{uhrich2023cavity}.

\begin{figure}[h]%
\centering
\includegraphics[width=0.5\textwidth]{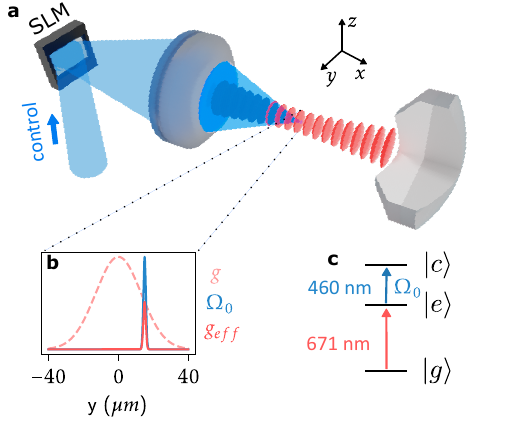}
\caption{\textbf{Concept of cavity-microscope. a} Overview of the system: a Fabry-Perot optical cavity is formed by two mirrors, which are optically contacted with a high aperture confocal lens pair sharing an optical axis with the cavity. The lens focuses a control beam (blue), for which the mirrors are transparent, onto a small volume within the cavity mode. A spatial light modulator controls the shape of this beam. 
\textbf{b} Spatial profile of the cavity mode (\protect\tikz[baseline]{\textcolor{WildStrawberry}{\protect\draw[line width=0.3mm] (0,.8ex)--++(.3,0) ;}}) and control beam (\protect\tikz[baseline]{\textcolor{NavyBlue}{\protect\draw[line width=0.3mm] (0,.8ex)--++(.3,0) ;}}). The control beam shapes the local light-matter coupling strength into $g_\text{eff}$ (\protect\tikz[baseline]{\textcolor{WildStrawberry}{\protect\draw[line width=0.3mm, densely dashed] (0,.8ex)--++(.3,0) ;}}), peaked at the control beam location (see text for details). 
\textbf{c} Level diagram of $^6$Li atomic transitions addressed by the cavity-microscope: the cavity has a high finesse around \SI{671}{\nano\meter}, close to the $\ket{2S}$ (labeled $\ket{g}$) to $\ket{2P}$ (labeled $\ket{e}$) transition. The control beam at \SI{460}{\nano\meter} couples states $\ket{2P}$ with $\ket{4D}$ (labeled $\ket{c}$) in a local fashion.
}\label{fig1}
\end{figure}

\begin{figure*}[ht]%
\centering
\includegraphics[width=0.9\textwidth]{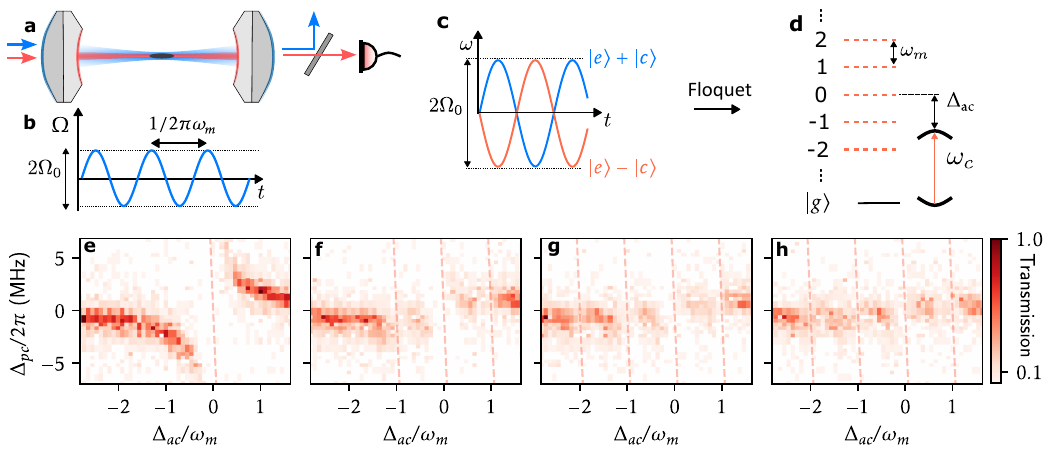}
\caption{\textbf{Floquet engineering of the light-matter coupling. a} The atomic cloud is homogeneously illuminated by the control beam while the cavity is interrogated with a probe beam at different frequencies. After separating the control beam and the cavity probe beam with a dichroic mirror, we record the transmission of the probe beam with a single photon counter. \textbf{b} The amplitude of the control beam is modulated in time at frequency $\omega_m$ such that the Rabi frequency, together with the instantaneous eigenenergies of the $\{\ket{e}$, $\ket{c}\}$ manifold and the corresponding eigenstates oscillate in time (\textbf{c}). This leads to a spectrum of equally spaced, spectrally isolated Floquet bands (\textbf{d}). Varying the cavity detuning with respect to the bare atomic transition ($\Delta_{ac}$), we observe strong light-matter coupling for $\Delta_{ac}\sim \omega_m$, with a coupling strength determined by the amplitude $\Omega_0$ of the control beam. \textbf{e-h} Cavity transmission spectra as a function of atom-cavity detuning and probe-cavity detuning ($\Delta_{pc}$) for different values of $\Omega_0$. At $\Omega_0=0$ (\textbf{e}), we see the avoided crossing when the cavity is on resonance with the $\vert g\rangle \longrightarrow \vert e\rangle$ state transition. As $\Omega_0/2\omega_m$ increases from $0$ to \qty{0.9} (\textbf{f}), \qty{1.2} (\textbf{g}) and \qty{1.8} (\textbf{h}) avoided crossings appear at multiples of the modulation frequency (marked in \protect\tikz[baseline]{\textcolor{WildStrawberry}{\protect\draw[line width=0.3mm, densely dashed] (0,.8ex)--++(.3,0) ;}}). The Rabi frequencies of the control beam are extracted by comparison with a direct numerical simulation (see Supplementary).
}\label{fig2}
\end{figure*}

\section{Results}
\subsection{Cavity-microscope device}\label{cavitymicroscope}

At the core of our system lies a new optical element, made of a high-reflectivity plane-concave cavity mirror optically contacted onto a custom-made plane-convex aspherical lens. The resulting element thus features a superpolished, high-reflectivity mirror then used to realize the cavity, while also operating as an aspherical lens for wavelengths transmitted by the mirror coating. A pair of those elements is then combined to form a near-concentric high-finesse cavity simultaneously with a confocal aspherical lens pair, with a common optical axis. The concept is presented in Fig. \ref{fig1}a.

Concretely, the cavity has a high finesse at \SI{671}{\nm} to near-resonantly address the D2 transition of $^6$Li, reaching a single-atom single-photon cooperativity of $\eta_0=$ \qty{2.55} for an atom trapped at the antinode of the cavity mode.The cavity is also resonant at \SI{1342}{\nm} for dipole trapping of the atoms (see Methods). The mirror-lens elements have an available numerical aperture (NA) of $0.35$ for atoms located at the center of the cavity. The mirrors are transparent at \SI{460}{\nm}, corresponding to the $2P\longrightarrow 4D$ transition in $^6$Li, such that light at this wavelength, focused by the lens-mirrors, can be used to dress the $2P$ state of the atoms locally as summarized in Fig. \ref{fig1}c \cite{Sauerwein_2023}. The local dressing of the $2P$ state imprints the geometry of the control beam on the spatial profile of the atom-cavity coupling, as shown in Fig.\ref{fig1}b. 

We combine the cavity-microscope with a Spatial Light Modulator (SLM) in the Fourier plane of one of the mirror-lens element, as shown in Fig. \ref{fig1}a, to allow for wavefront engineering. This is optimal for atoms tightly confined by a cavity-enhanced optical lattice into a quasi-two dimensional geometry, with the highest resolution in the plane perpendicular to the optical axis. It is compatible with the tweezer-cavity architecture for quantum information processing, since atoms can be moved in and out of the cavity mode along the lateral directions. 

\subsection{Floquet engineering of the light-matter coupling}\label{floquet}

To control the atom-cavity coupling, we employ a Floquet engineering approach \cite{Goldman:2014ac,Eckardt:2015aa,Bukov:2015ab,clark2018interacting}, where the $2P\longrightarrow 4D$ transition is resonantly driven with a laser beam propagating through the cavity-lens system, as shown in Fig. \ref{fig2}a. This light field is then amplitude modulated with frequency $\omega_m$ (Fig. \ref{fig2}b), yielding a time-dependent Rabi frequency $\Omega(\mathbf{r},t) = \Omega_0(\mathbf{r}) \cos \omega_m t$. The position dependence of the Rabi frequency is due to the spatial profile of the control laser. The instantaneous eigenfrequencies and eigenstates, corresponding to symmetric and anti-symmetric combinations of the driven $2P$--$4D$ system, thus oscillate periodically in time as illustrated in Fig. \ref{fig2}c. The Fourier components of these eigenenergies form Floquet bands separated by $\omega_m$ that replace the bare optical spectrum of atoms (Fig. \ref{fig2}d). The spectral weight of the different bands inherits a space dependence from the position dependence in $\Omega_0(\mathbf{r})$: where $\Omega_0 = 0$ all but the zeroth order band vanish and the bare spectrum is recovered; elsewhere, each band behaves as a spectrally isolated atomic resonance with an atom-cavity coupling determined by the local values of $\Omega_0$ (see Methods for details). 


We demonstrate Floquet engineering by performing transmission spectroscopy of the cavity hosting the atoms exposed to the control beam, as a function of both the atom-cavity and probe-cavity detunings, $\Delta_{ac}$ and $\Delta_{pc}$ respectively. We first explore the simple case where all the atoms are homogeneously addressed by the control beam. This is done by increasing the waist of the control beam at focus using a small numerical aperture on the cavity lens. We choose $\omega_m/2\pi=$ \SI{110}{\MHz}, large compared to the collective atom-cavity coupling $g\sqrt{N}/2\pi = \SI{16}{\MHz}$, to allow for independent addressing of the different bands with the cavity. In the absence of dressing, we observe the expected vacuum Rabi splitting (Fig. \ref{fig2}e). As the control laser intensity is increased, the bare avoided crossing pattern is reduced and smaller avoided crossings emerge at $\Delta_{ac}=\pm\omega_m$, as shown in Fig. \ref{fig2}f. For even larger control beam intensities (Fig. \ref{fig2}g and h), transmission drops are observed also for higher order Floquet bands $\Delta_{ac}=\pm2\omega_m$, as expected in the regime $\Omega_0\sim2\omega_m$. 

To interpret our observations with a simple model, we model the transmission of the cavity close to a Floquet band using an effective cooperativity, capturing the spectral weight transferred to a given band by the modulation. For instance, on resonance with the $n^{th}$ band, the transmission of the cavity in the presence of the dressed atoms can be expressed  as $1/(1+\eta^{(n)}_\mathrm{eff})^2$, with an effective cooperativity 
\begin{equation}
\eta^{(n)}_\mathrm{eff} = \int \rho(\mathbf{r})\frac{4g_n^2(\mathbf{r})}{\kappa\Gamma} d^3\mathbf{r}
\label{eq:effectiveCooperativity}
\end{equation}
where $\rho(\mathbf{r})$ is the atomic density. The coupling strengths $g_n(\mathbf{r}) = g(\mathbf{r}) \, J_n\left(\frac{\Omega_0(\mathbf{r})}{2\omega_m}\right)$, where $g(\mathbf{r})$ is the shape of the cavity mode, are given by the Bessel function of order $n$ (see \cite{clark2018interacting}). Importantly, $g_n$ features a position dependence via its non-linear relation with the control beam Rabi frequency $\Omega_0(\mathbf{r})$. This effective description is benchmarked against a direct numerical simulation of the system in Fig. \ref{figM1}a-b and its range of validity is discussed in Methods.

Floquet engineering offers several decisive advantages compared to other techniques for spectral tuning of atomic transitions. For instance, in contrast with light shifting methods \cite{Brantut:2008aa,veyron:2023aa,Sauerwein_2023,Baghdad_2023}, it yields discrete resonances at frequencies dependent only on the modulation frequency, making it robust against intensity or beam shape fluctuations. Far-detuned two-photon transitions using a red and a blue photon would also allow for a similar addressing without the drawbacks of light shifts, however with a significant decrease of the light-matter coupling. The lambda-level scheme proposed in \cite{Yang:2018aa,Yang:2018ab} would allow for similar capabilities, but requires two-electron atoms with optical clock transitions, or atoms with large hyperfine splittings where the control beam and the cavity have to be sufficiently detuned from each other in order not to lead to spurious couplings. Compared with \cite{clark2018interacting}, where Floquet engineering was achieved using the modulation of a light shift, our resonant drive with amplitude modulation offers the largest frequency modulation amplitudes, allowing to operate with large values of $\omega_m$.

\begin{figure}[h]%
\centering
\includegraphics[width=0.5\textwidth]{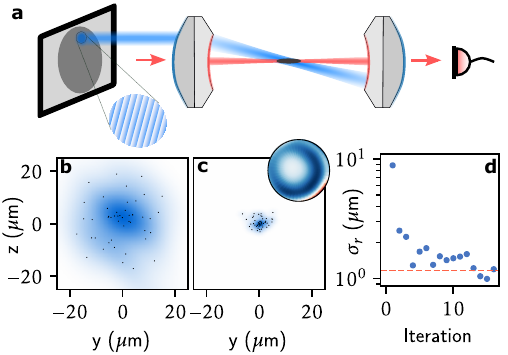}
\caption{\textbf{Cavity-assisted wavefront sensing and correction. a} The SLM surface is divided in patches and rays emerging from one patch at the time are directed on the atoms. We measure cavity transmission for different ray directions for a given patch and search for maximum atom losses to align the ray onto the atoms. \textbf{b,c} Dispersion of rays in the plane of the atoms: each point corresponds to rays from a given patch, before (b) and after (c) applying a wavefront correction. Inset: optimal wavefront correction \textbf{d} Root mean square deviation of the pointing across the wavefront, as a function of iteration of the correction algorithm. We limit the aperture to NA $=$ \qty{0.24}, for which the RMS deviation of the pointing reaches the diffraction limit (\protect\tikz[baseline]{\textcolor{WildStrawberry}{\protect\draw[line width=0.3mm, densely dashed] (0,.8ex)--++(.3,0) ;}}).}
\label{fig3}
\end{figure}

\subsection{Wavefront engineering}\label{aberrationcorrection}

The Floquet dressing, as shown in equation \ref{eq:effectiveCooperativity}, allows to turn atoms in the cavity into a high-sensitivity sensor for the intensity of the control beam. Indeed, cavity-assisted absorption spectroscopy resonant with one of the Floquet bands directly provides information about the intensity of the control beam. We now use this feature to optimise the wavefront of the control beam at the location of the atoms, similar to ~\cite{Zupancic:2016ab, PhysRevX.4.021034}. 

Our procedure is a variant of the Shack-Hartmann method. We prepare a small atomic cloud trapped at the center of the cavity mode, then use the spatial light modulator to select a small patch of the wavefront and direct the corresponding light pencil onto the atomic cloud (Fig. \ref{fig3}a). We set the cavity on resonance with the first Floquet band, and probe the cavity on-axis using light resonant with the bare cavity resonance. We measure atom losses, induced by the on-axis probe, which shows a higher signal to noise ratio than photon transmission for this application. We then repeat this measurement and steer the light pencil using the SLM to search for the loss maximum (see Supplementary), signaling optimal alignment of the pencil on the atoms. After collecting this information for all patches of the wavefront, we obtain a ray-tracing diagram of the type of Fig. \ref{fig3}b-c, and infer the overall wavefront shape (see Methods, Fig. \ref{figM2}a-c). An iterative algorithm is then used to correct the measured aberrations and optimize the beam quality directly on the atoms. 

The results of the optimization are presented in Fig. \ref{fig3}d. As a figure of merit we choose the root-mean-square deviation of the ray-tracing measurement which is directly measured in the experiment. It converges after about $6$ iterations, and consistently reaches below \SI{1}{\micro\meter} after $12$ iterations of the algorithm. The resulting ray distribution and deduced wavefront imperfections are presented in Fig. \ref{fig3}c. After the wavefront correction, we choose to operate with a numerical aperture of \qty{0.24}, which corresponds to a diffraction limit of $1.22\lambda/2$NA$ = $\SI{1.17}{\micro\meter}. We observe a RMS deviation of the rays below this value (see Fig.\ref{fig3}d), for which we estimate a Strehl ratio of \qty{0.67}, approaching the diffraction limit. This is below the full available NA of the cavity lens, mainly due to the reduced resolution of the ray-tracing measurement procedure for parts of the wavefront that are further away from the optical axis (see Supplementary). Indeed, the finite extension of the cloud along the optical axis renders accurate measurements of the beam pointing more difficult. With this restriction, we obtain a control beam with size one order of magnitude smaller than the cavity mode waist (see Methods for details).

This result also informs about the optical quality of the cavity-lens system as such. The main contributions to the large initial wavefront distortion are defocus and spherical aberrations (see Methods, Fig. \ref{figM2}d). Indeed, having in mind future applications of this system in quantum simulations with small clouds or single atoms \cite{Serwane_2011,spar:2022aa}, the cavity-lens has been designed for optimal performance as an optical tweezer trap at \SI{780}{\nm}, in addition to the \SI{460}{\nm} dressing. Its aspherical surface was not expected to yield a flat wavefront at \SI{460}{\nm}. Large coma is also observed, which we attribute to the imperfect alignment of the lens and cavity optical axis. Remarkably, the cavity-assisted wavefront sensing and engineering allows to significantly suppress these imperfections. 

\subsection{Scanning probe measurement}\label{tomography}

\begin{figure}
\centering
\includegraphics[width=0.5\textwidth]{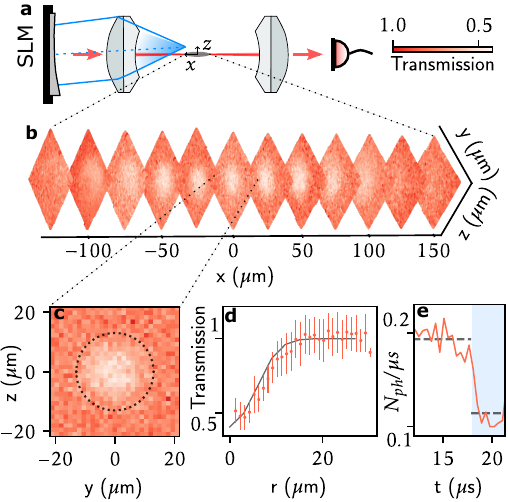}
\caption{\textbf{Scanning probe measurement of the atomic cloud. a} Scanning of the position of the focus of the control beam using the SLM. For each position, the cavity transmission is measured for the cavity tuned near the second Floquet band. \textbf{b} Scanning probe images of a thermal cloud trapped in the cavity mode, for different locations $x$ along the cavity direction. Each pixel at location $y,z$ in a given picture represents the average of \qty{25} cavity transmission measurements for the control beam focused at location $y,z$. \textbf{c} Image of the central part of the cloud.  Dashed, the waist of the cavity field at the center of the cavity. \textbf{d} Comparison of the azimuthally averaged cavity transmission measurements for $x$= \qty{0} (\textcolor{WildStrawberry}{$\bullet$}) with the direct numerical simulation of the system (\protect\tikz[baseline]{\textcolor{Gray}{\protect\draw[line width=0.3mm] (0,.8ex)--++(.3,0) ;}}), peaked at the control beam location (see text for details) for a maximal control Rabi frequency of $\Omega_0/2\omega_m=$ \qty{5.2}. \textbf{e} Fast, local measurements: transmission of the cavity upon instantaneous turn-on of the control beam (at time $t=$ \SI{18}{\micro\second}, blue background), when the control beam is focused at the center of the cloud. Dashed, the average photon count rate when the control beam is off and on.}\label{fig4}
\end{figure}

With a tightly focused control beam, we now demonstrate the operation of our system as a cavity-assisted scanning probe microscope with micrometer-scale resolution \cite{Yang:2018aa,Yang:2018ab}. Inspired by other scanning probe microscopy techniques \cite{Mader:2015aa, Gericke:2008aa, veit:2021aa}, we use the focus of the control beam to single-out atoms in a microscopic region of space and perform cavity-assisted spectroscopy on those. Scanning the position of the focal spot across the volume of the cavity, we reconstruct the atomic density distribution. Specifically, we tune the cavity resonance \SI{-8}{\mega\Hz} away from the second Floquet band and the cavity-probe detuning $\Delta_{pc} =$ \SI{2}{\mega\Hz} on the dispersively shifted cavity resonance. We found that this choice of Floquet band mitigates residual off-resonant absorption due to atoms out of the control beam, while keeping a sufficiently high effective cooperativity. 
We then use the SLM to set the position of the focus and perform transmission spectroscopy for \SI{20}{\micro\s}, as illustrated in Fig. \ref{fig4}a. Collecting transmission signals recorded for spot locations spanning a two-dimensional region yields a scanning probe image of the atoms. Figure \ref{fig4}b presents a set of two-dimensional images recorded with this technique, with the focus of the control beam varied through the cloud along the cavity axis direction.  Figure \ref{fig4}c presents the details of such an image. Remarkably, the measurement of the local atomic density is obtained purely through cavity-assisted measurements, even though the entire cloud is contained in the mode of the cavity. 


We compare the profile produced by the scanning probe technique with the cavity transmission obtained from a direct numerical simulation of the system, presented in Fig. \ref{fig4}d with a solid line. The direct simulation takes into account the temperature of the cloud, trap geometry, atom-cavity and probe-cavity detunings, total atom number and NA of the lens. These parameters determine the spatial inhomogeneity of the cloud, cavity mode and control light. As explained in Methods \ref{M:direct_simulation}, the simulation takes into account the effect of all Floquet bands up to order \qty{12}. Leaving the peak control beam Rabi frequency $\Omega_0$ as a free parameter, we obtain excellent agreement with the experiment for $\Omega_0/2\omega_m =  $\qty{5.2}, validating our scanning probe method for local measurements of the atomic density.
The data also agree qualitatively with the effective cooperativity model, up to corrections due to higher order Floquet bands (see Methods for details). \\


Last, in addition to the spatial resolution, our scanning probe microscope also offers time resolution at very short scales. Switching on and off the control beam, we can determine the atomic density at the location of the control beam focus within microseconds using the cavity. This is illustrated in Fig. \ref{fig4}e, where we present the transmitted photon count rate as the control beam is switched on, yielding a time-scale for the transmission drop at the microsecond scale, ultimately limited by the light-matter interaction strength, while retaining the spatial resolution permitted by the tightly focused control beam. 

\section{Discussion}

Our results demonstrate that the space dependence of the atom-photon couplings in cavity QED does not need to be constrained by the mode structure inherited from the geometry of the cavity, but can instead be shaped in space and time in a programmable fashion. Our cavity-microscope device makes use of this capability in a single optical device, making it both simple and robust. For instance, our wavefront sensing and shaping protocol needs to be done only once, being immune to relative misalignments of the atomic cloud, cavity and high-aperture lens. 
In our simple approach with amplitude modulated resonant dressing, the effective cooperativity is reduced compared with the bare one. With more laser power available for dressing, this could be circumvented using a light shift modulation rather than a modulated resonant drive \cite{clark2018interacting}.  This would limit the admixture of the Floquet bands with state $\ket{c}$, which is not coupled to the cavity, thus maximizing the atom-cavity coupling. Furthermore, serrodyne rather than sinusoidal modulation, comprising higher order harmonics, would allow to transfer all the spectral weight into one chosen target band and yield an effective cooperativity equal to the bare one.

Because of the non-linear relation between the local atom-cavity coupling and the local control field strength, the spatial resolution of the scanning probe measurement is not bound by the diffraction limit of the cavity lens. In fact, we can expect super-resolution, with an improvement by a factor about $\sqrt{n}$ compared to the diffraction limit for probing on a Floquet band of order $n$. For Lithium atoms trapped in a typical tweezer with trap frequency of \SI{10}{\kHz} \cite{Serwane_2011}, the length scale of the ground state wavefunction is \SI{410}{\nm}, within the reach of our probing technique. 

The cavity-microscope can also be operated in the context of quantum information applications, where the cavity-mediated flip-flop interactions enable all-to-all couplings between atomic qubits \cite{sorensen:2003aa,borregaard:2015aa,zheng:2000aa,pellizzari:1995aa}. Combined with fast programmable switching of the interactions, this allows for full-range multi-qubit gates. Furthermore, the fast local addressability allows for mid-circuit readout of the atomic qubits, eliminating the need to move the atoms in the cavity field. 

Last, our system is ideally suited for combining this fast local control over light-matter interactions with the preparation of small ultra-cold samples of fermionic $^6$Li atoms. Indeed, the cavity lenses have been designed also for producing tight optical tweezers using light at \SI{780}{\nm}, such that the method of \cite{Serwane_2011,spar:2022aa} could be directly implemented in the cavity. This will enable the study of fermionic systems with strong and programmable long-range interactions \cite{helson:2023aa}. Such a capability would allow to turn the cavity-mediated interactions into the fully-random, all-to-all coupling necessary for the quantum simulation of holographic quantum matter such as the Sachdev-Ye-Kitaev model \cite{chowdhury:2022aa,uhrich2023cavity}.

 \clearpage

\section{Methods}\label{methods}

\subsection{Theoretical description}

\subsubsection{Effective cooperativity}

Here we compute the effective light-matter coupling in the presence of the Floquet dressing. We focus on the regime of weak cavity probe where the photon number is very low, and restrict the description to the single excitation regime. The Hamiltonian of the system can be written as
\begin{equation}
\hat{H} = \hat{H}_0 + \hat{H}_\text{int} 
\end{equation}
with 
\begin{multline}
\hat{H}_0 = \sum_j \omega_e | e \rangle_j \langle e |_j + \omega_c | c \rangle_j \langle c |_j + \omega_{\mathbb{C}} \hat a^\dagger \hat a +
\\ \Omega_j \cos(\omega_m t)\cos\left[(\omega_e - \omega_c)t\right]\left( | c \rangle_j \langle e |_j + | e \rangle_j \langle c |_j\right)
\end{multline}

where $\omega_{e,c}$ designate the energies of the $| 2P \rangle$ and $| 4D \rangle$ states respectively. The index $j$ runs over the $N$ atoms of the gas, with $\Omega_j$ the local Rabi frequency of the control beam, and states $ | e,c \rangle_j$ represent the state in which all atoms except $j$ are in the ground state, and atom $j$ is in state $e$ or $c$. $\hat a$ annihilates photons in the cavity, and $\omega_{\mathbb{C}}$ is the cavity resonance frequency. $\omega_m$ is the modulation frequency of the amplitude of the control beam. 

The atoms-photon interactions is
\begin{equation}
\hat{H}_\text{int} = \sum_j g \left( \hat a ^\dagger| g,g...g \rangle \langle e |_j + \hat a| e \rangle_j \langle g,g...g | \right),
\end{equation}
which can be rewritten in terms of the symmetric states 
\begin{equation}
|W_e\rangle = \frac{1}{\sqrt{N}} \sum_j | e \rangle_j ,\, |G\rangle = \hat{a}^\dagger |g,g,...,g\rangle
\end{equation}
yielding the familiar
\begin{equation}
\hat{H}_\text{int} = g \sqrt{N}\left( |G\rangle \langle W_e| +hc \right).
\end{equation}

For a resonant drive, the dressing part of the Hamiltonian is diagonal when written in terms of the  states $|\pm\rangle_j = \frac{1}{\sqrt{2}}(|e\rangle_j \pm |c\rangle_j)$. In the frame rotating at the atomic transition frequencies, 
\begin{multline}
\hat{H} = \sum_j \frac{\Omega_j}{2} \cos(\omega_m t)\left( | + \rangle_j \langle + |_j - | - \rangle_j\langle -|_j\right) \\+ g \sqrt{N}\left( |G\rangle \langle W_e| +hc \right) + (\omega_{\mathbb{C}} - \omega_e) |G\rangle \langle G|.
\end{multline}

To make the Floquet band spectrum explicit, we introduce the time-dependent unitary transformation acting on the atomic excited states manifold, to eliminate the dressing:
\begin{multline}
\hat{U}(t) = \text{exp}\left[ \frac{i \sin(\omega_m t)}{2\omega_m}  \sum_j \Omega_j (| + \rangle_j \langle + |_j - | - \rangle_j\langle -|_j)\right]\\
= \sum_{j,n} J_n\left( \frac{\Omega_j}{2\omega_m} \right) e^{ i n \omega_m t}\left[ | + \rangle_j \langle + |_j 
+ (-1)^n  | - \rangle_j\langle -|_j \right]
\end{multline}
where the second line follows from the Jacobi-Anger expansion, and $J_n$ are the $n$th order Bessel functions. 

The state coupled to light then breaks into a set of harmonics:
\begin{multline}
\hat{U}(t) |W_e\rangle \\= \frac{1}{\sqrt{2N}} \sum_{n,j} J_n\left( \frac{\Omega_j}{2\omega_m} \right) e^{ i n \omega_m t} \left[ | + \rangle_j + (-1)^n  | - \rangle_j \right] \\ = \frac{1}{\sqrt{2N}} \sum_{n} e^{ i n \omega_m t} \sqrt{A_{n}}\left( |W_{+,n} \rangle + |W_{-,n} \rangle \right) 
\end{multline}

Each harmonic in this expression represents the particular single-excitation state coupled to the cavity photons in the corresponding Floquet band:
\begin{equation}
|W_{\pm,n}\rangle =\frac{(\pm1)^n }{\sqrt{A_{n}}} \sum_j J_n\left( \frac{\Omega_j}{2\omega_m} \right) | \pm \rangle_j
\end{equation}
with a normalization factor $A_{n} = \sum_j J_n^2\left( \frac{\Omega_j}{2\omega_m} \right)$ representing the effective atom number participating into a particular Floquet band-cavity coupling. 

After unitary transformation, the Hamiltonian is written as
\begin{equation}
\hat{H} =  (\omega_{\mathbb{C}} - \omega_e) |G\rangle \langle G| + \sum_{n\in\mathbb{Z}} \hat{H}_n e^{ i n \omega_m t} 
\end{equation}
and 
\begin{equation}
\hat{H}_n =  g \sqrt{A_n}  |G\rangle \left( \langle W_{+,n}| + \langle W_{-,n}|  \right) +hc
\end{equation}
For a cavity resonant with one of the Floquet band $\omega_{\mathbb{C}} = \omega_e +n\omega_m$, in the rotating wave approximation with respect to the modulation frequency (or equivalently the zeroth order in the Magnus expansion), the light-matter coupling Hamiltonian reduces to $\hat{H}_n $. The effective, many-atoms cooperativity is then 
\begin{multline}
\eta_\text{eff}^{(n)} \hat{=} \frac{4g^2A_n}{\kappa \Gamma}  = \frac{4g^2}{\kappa \Gamma} \sum_j J_n^2\left( \frac{\Omega_j}{2\omega_m} \right) \\= \int \rho(\mathbf{r}) \frac{4g^2}{\kappa \Gamma} J_n^2\left( \frac{\Omega(\mathbf{r})}{2\omega_m} \right) \;  d\mathbf{r}, 
\end{multline}
as in equation \ref{eq:effectiveCooperativity}.

In practice, for calculations we also include the spatial variations of the cavity mode $g(\mathbf{r})$ and use the measured temperature, atom number and trap parameters to predict $\rho(\mathbf{r})$. Consistent with the assumptions of this treatment, we then treat the Floquet band as an isolated resonance and predict the transmission for the $n^\text{th}$ band using \cite{tanjisuzuki2011interaction}
\begin{equation}\label{eq:effective_cooperativity_simulation}
    T^{(n)} = \frac{1}{(1+\frac{\Gamma_e^2}{\Gamma_e^2+4\Delta_{ac}^2}\eta_{eff}^n)^2 + (\frac{-2\Delta_{ac}\Gamma_e}{\Gamma_e^2+4\Delta_{ac}^2}\eta_{eff}^n)^2}
\end{equation}



This approach neglects the dispersive shifts and residual absorption from the off-resonant Floquet bands, that enter in higher-order terms in the Floquet-Magnus expansion. 


\subsubsection{Direct numerical simulation}\label{M:direct_simulation}

In order to go beyond the intuitive, low-order treatment above yielding the effective coooperativity and account for higher-order processes, we turn to a direct numerical calculation of the transmission spectrum. 
We start from the Hamiltonian in the single excitation manifold, written in the frame rotating at the dressing laser frequency, in the resonant dressing regime. For a homogeneous dressing beam, we obtain:
\begin{multline}
\hat{H} = \omega_e \left( | W_e \rangle \langle W_e | + | W_c \rangle \langle W_c | \right) +\\ 
 \frac \Omega 2 \cos(\omega_m t)\left( | W_c \rangle \langle W_e | + | W_e \rangle \langle W_c |\right) + \\ \omega_{\mathbb{C}} | G \rangle \langle G | + g \sqrt{N} \left(| G \rangle \langle W_e | + | W_e \rangle \langle G |\right),
\end{multline}
where 
\begin{equation}
|W_c\rangle = \frac{1}{\sqrt{N}} \sum_j | c \rangle_j .
\end{equation}

Following \cite{Eckardt:2015aa}, we then introduce a time-independent quasi-energy operator $\bar{Q}$ acting on an extended Hilbert space $ \bar{H} = \text{Sp}\left\{ | G \rangle, | W_e \rangle, | W_c \rangle\right\} \otimes \mathcal{F}_\omega$, where $\mathcal{F}_\omega$ is the Hilbert space of $2\pi/\omega-$periodic functions, spanned by all the harmonics of the modulation frequency. To account for the first $\mathcal{N}$ Floquet bands, we truncate $\mathcal{F}_\omega$ to a dimension $\mathcal{N}$, leading to a $3(2\mathcal{N}+1) \times 3(2\mathcal{N}+1)$ quasi-energy matrix.

To obtain the transmission of the system to first order in the on-axis cavity drive, we evaluate 
\begin{equation}\label{eq:transmission}
T \propto |\langle G |  \frac{1}{- \bar{Q}} | G \rangle|^2
\end{equation}
by directly inverting the matrix. The finite decay rate of the photons and of the excited states is accounted for by introducing a corresponding imaginary part to the bare $\omega_{\mathbb{C},e,c}$ before the inversion. 

In order to account for the spatial inhomogeneity of the atomic cloud, the mode function of the cavity and control beam, we consider the small volume around location $x$, containing $N_x$ atoms as a homogeneous ensemble coupled to the cavity and described by the formalism above. The full system then obeys  
\begin{multline}
\hat{H} = \\  \omega_{\mathbb{C}}| G \rangle \langle G |  + \sum_x g \sqrt{N_x} \left(| G \rangle \langle W_{e,x} | + | W_{e,x} \rangle \langle G |\right)+ \\
 \omega_e \left( | W_{e,x} \rangle \langle W_{e,x} | + | W_{c,x} \rangle \langle W_{c,x} | \right) + \\
\frac{\Omega_x}{2} \cos(\omega t)\left( | W_{c,x} \rangle \langle W_{e,x} | + | W_{e,x} \rangle \langle W_{c,x} |\right) 
\end{multline}
where now all the couplings and atom numbers acquire a label $x$ describing their local values. With $N = \sum_x  N_x $ this reproduces the all-to-all coupling induced by the cavity

\begin{multline}
\sum_x g \sqrt{N_x} \left(| G \rangle \langle W_{e,x} | + | W_{e,x} \rangle \langle G |\right) \\= g\sqrt{N} \left(| G \rangle \langle W_{e} | + | W_{e} \rangle \langle G |\right)
\end{multline}

Dividing space into $M$ such small volumes yields an overall quasi-energy matrix of dimension $(2M+1)(2\mathcal{N}+1) \times (2M+1)(2\mathcal{N}+1)$ which we invert numerically to obtain the simulations presented in Fig. \ref{figS1}e-h. 


To compare with the experiments, we introduce an $x$-dependent coupling strength $g_x$ to describe the mode profile. We account for the finite lifetime of the photons in the cavity and the excited atomic states using a corresponding imaginary part to the atomic and cavity energies, which is accurate for linear response calculations \cite{clark2018interacting}. 



In practice, we divide the space in sectors in the three-dimensional space, with resolutions $\Delta_y=\Delta_z =$ \SI{1}{\micro\m} and $\Delta_x =$ \SI{10}{\micro\m} along the transverse and longitudinal directions, respectively, and average the profiles of the mode shape, atomic cloud and control beam in each volume. This is chosen to keep the dimensions of the quasi-energy matrix to $25000\times25000$, which we directly invert to obtain the transmission. Iterating this procedure for each location of the focus of the control beam, we obtain the simulation shown in Fig. \ref{fig4}d.\\

\begin{figure}[h] 
    \centering
    \includegraphics[width=0.5\textwidth]{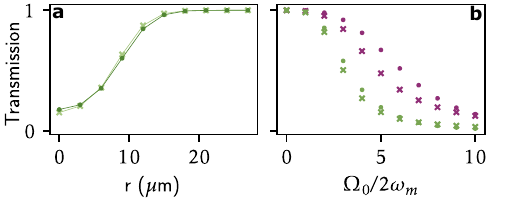}
    \caption{\textbf{Comparison between direct numerical simulation and effective cooperativity formula. }\textbf{a} Cavity transmission as a function of control beam position, calculated with the direct numerical simulation ($\times$) and the effective cooperativity formula of Eq. \ref{eq:effective_cooperativity_simulation} ($\bullet$), for the cavity on resonance with
    the second Floquet band ($\Delta_{ac}= 2\omega_m =$ \SI{220}{\mega\Hz}). \textbf{b} Cavity transmission with the control beam focused at the center of the cloud ($r=$ \SI{0}{\micro\m}) for different control beam powers and detunings from the second Floquet band: $\Delta_{ac}=$ \SI{220}{\mega\Hz} (\textcolor{ForestGreen}{$\bullet, \times$}) and  $\Delta_{ac}=$ \SI{212}{\mega\Hz} (\textcolor{Purple}{$\bullet, \times$}), which is the experimental configuration of Fig. \ref{fig4}b-d.} \label{figM1}
\end{figure}

We compare the cavity transmission obtained from the direct numerical simulations and from the effective cooperativity model for different maximum Rabi frequencies of the control beam and different atom-cavity detunings, on and off-resonance with the second Floquet band.
Figure \ref{figM1}a presents the cavity transmission profile for the scanning probe microscope measurement of Fig. \ref{fig4}b-c, for the cavity on resonance with the second Floquet band and for a maximum Rabi frequency of $\Omega_0/2\omega_m =$ \qty{5.2}. The effective cooperativity formula of Eq. \ref{eq:effectiveCooperativity} quantitatively captures the cavity-enhanced absorption for every position of the control beam focus. Figure \ref{figM1}b shows the limitations of the effective cooperativity formula for the off-resonant case of Eq. \ref{eq:effective_cooperativity_simulation}. For the control beam focused at the center of the cavity ($r=$ \qty{0}), we calculate the cavity transmission for different values of the maximum Rabi frequency of the control beam. While the cavity-enhanced absorption is correctly captured when the cavity is on resonance with the second Floquet band ($\Delta_{ac}= 2\omega_m$), we observe deviations when moving away from the resonance, in particular for $\Delta_{ac} - 2\omega_m = $ \SI{-8}{\mega\Hz}. We attribute the discrepancy to higher order terms in the Floquet-Magnus expansion, such as dispersive coupling or off-resonant absorption from higher Floquet bands, neglected in the effective cooperativity model.



\subsection{Cavity-Microscope parameters}


\subsubsection{Cavity mirrors}
The mirrors have a diameter of \SI{16}{\mm} and a clear aperture of \SI{10}{\mm}. The radius of curvature is \SI{13}{\mm} and the cavity length is \SI{25.9}{\mm}, i.e. the cavity operates about \SI{100}{\micro\m} from the concentric limit, which yields a Rayleigh range of $Z_\mathrm{R}$= \SI{850}{\micro\metre}. This geometry was chosen to maximise the single-atom single-photon coupling \cite{PhysRevA.96.031802,doi:10.1126/science.1219166, PhysRevLett.122.010405}. The free spectral range of the cavity is $\nu_\mathrm{FSR}=\SI{5.77}{\GHz}$ and the transverse mode spacing is $\nu_\mathrm{TMS}=\SI{240}{\MHz}$.

The finesse of the cavity at \SI{671}{\nm} is \num{5.2e3} and the beam waist at the center is \SI{13}{\micro\metre}, deduced from the transverse mode spacing, yielding a cooperativity of \num{2.52} and a single-atom single-photon coupling of $g / 2\pi = $\SI{2}{\MHz} for the cycling transition. The cavity is also resonant at \SI{1342}{\nm}, allowing to produce a deep intra-cavity dipole trap with a mode waist of \SI{19}{\micro\metre}. Light resonant with the atoms is produced by second harmonic generation from a \SI{1342}{\nm} beam stabilised on the fourth-order transverse mode of the cavity. The mirrors coating also feature high transmission at \SI{460}{\nm} and at \SI{780}{\nm}.\\

\subsubsection{Cavity lenses}

The aspherical lenses contacted onto the mirrors have a diameter of \SI{16}{\mm} and a clear aperture of \SI{14.4}{\mm}, offering a maximum numerical aperture of NA$=$ \num{0.35}. The lenses aspherical coefficients and material were optimized for performance at \SI{780}{\nm} for a working distance equal to half the mirror spacing, anticipating a future operation of the cavity-microscope as an optical tweezer. Operating at \SI{460}{\nm} leads to a chromatic shift of about \SI{1.4}{\mm}. We correct for this by using a diverging beam incident on the cavity lens, and rely on the SLM for finer adjustments of the focus. The mirror-lens elements rest on shear-piezoelectric actuators for active stabilization of the cavity length, and both are glued on a vibration damping platform \cite{Sauerwein_2022}. 


\subsection{Experimental procedure}


\subsubsection{Atoms preparation}\label{M:atoms_preparation}

The preparation of the atomic sample is identical to that employed in our previous work \cite{Sauerwein_2023}. In short, we prepare a thermal cloud of $^6$Li atoms trapped in a cavity-enhanced dipole trap at \SI{1342}{\nm}, directly loaded at the center of the cavity from a magneto-optical trap. In the longitudinal direction, the trapping beam is a standing wave and the cloud is made up of individual pancake-shaped trap sites spaced by \SI{671}{\nm}. The geometry of the cavity ensures maximum overlap between the trap sites and the antinodes of the cavity mode at \SI{671}{\nm}. From time-of-flight measurements, we estimate a temperature of the atomic cloud of \SI{480}{\micro\kelvin}. We measured trap frequencies of \SI{17.5}{\kHz} in the transverse direction and \SI{1.7}{\MHz} along the cavity direction from modulation spectroscopy. This gives us a thermal radius of the cloud of \SI{7.4}{\micro\meter}.  

We empty all but the central $180$ sites of the cavity-dipole trap using radiation pressure, by imaging an opaque mask on the center of the cloud with a laser resonant on the D2 transitions \cite{Sauerwein_2023}. We then repump the atoms in the hyperfine manifold $F=1/2$ of the D2 transition with an on-resonant cooler beam. 

\subsubsection{Dressing and interrogation}

The control beam at \SI{460}{\nm} is stabilized on resonance with the $2P_{3/2}\longrightarrow 4D_{5/2}$ transition of lithium using a high-accuracy wavemeter, also compensating for a trap-induced light shift of about \SI{200}{\mega\Hz} for this transition. Amplitude modulation is then introduced by combining one positive and one negative order of diffraction from an acousto-optic modulator. 

The interrogation of the atoms in the cavity then consists in a simultaneous pulse of the control and cavity probe beams. We use pulse durations of \SI{20}{\micro\s} for the scanning probe measurements (Fig. \ref{fig4}b-c), and up to \SI{200}{\micro\s} for the Floquet avoided crossing measurements (Fig. \ref{fig2}e-f, and transmitted light is collected on a single photon counter. Before and after interrogation, we count the total number of atoms via the dispersive shift of the cavity (without any control light), allowing to infer losses for each individual realization.

\begin{figure*}[h]%
    \centering
    \includegraphics[width=0.9\textwidth]{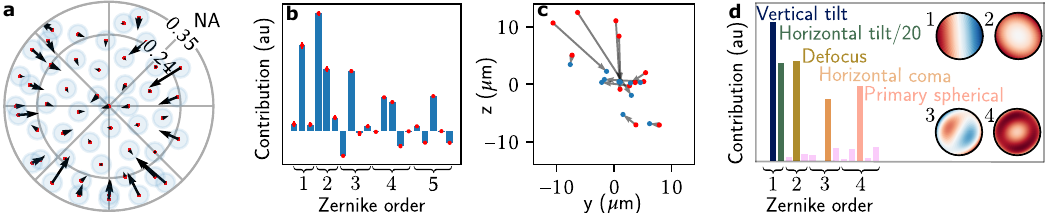}
    \caption{\textbf{Wavefront analysis and correction.}  \textbf{a} Typical raw data for a wavefront analysis: for each patch of the control beam (blue circles), we estimate the local phase gradient needed to reach the atomic cloud in the center, indicated by the direction and length of the arrows. \textbf{b} Fit results of the data of \textbf{a} across the wavefront with Zernike polynomials, yielding the relative contribution of the different orders. \textbf{c} Expected improvement of the rays pointing on the atom plane after applying the wavefront correction on the SLM. For 13 patches of the wavefront, we show the pointings before  (\textcolor{WildStrawberry}{$\bullet$}) and after (\textcolor{NavyBlue}{$\bullet$}) the correction . This correction improves the RMS deviation of the pointings of $\sigma_r =$ \SI{5}{\micro\meter}  to  $\sigma_r =$ \SI{2.7}{\micro\meter}. \textbf{d} Final wavefront correction applied on the SLM after 17 iterations. The spherical aberration and horizontal coma are dominant.  Inset: applied corrections of the wavefront at different Zernike orders.}\label{figM2}
\end{figure*}


\subsection{Wavefront engineering} \label{M:wavefront_engineering}
\subsubsection{In-situ wavefront correction}
Our wavefront correction measurement is inspired by the Hartmann-Shack procedure. We divide the wavefront on the SLM into circular patches of equal diameter, and sequentially characterize the wavefront orientation within each patch, as shown in Fig. \ref{figM2}a. The dimension of each patch represents about a tenth of the whole beam diameter, and was chosen as a compromise between resolution and signal-to-noise ratio.

We select a given patch by applying a phase gradient over the whole SLM area except on the patch area. This sends light from all the wavefront away from the atoms, except the light of the chosen patch. 

Applying a phase gradient along the $(y,z)$ directions steers the ray originating from this patch in the focus plane. Using an optimisation algorithm described in Supplementary, we find the optimal phase gradient for the patch by maximizing atom losses.\\

Once all patches have been measured, we have obtained an estimate the phase gradient in the entire beam area. An example of this measurement is shown in Fig. \ref{figM2}a for the iteration number \qty{9} of Fig. \ref{fig3}d. We then reconstruct the wavefront correction pattern by breaking down the global phase pattern into Zernike polynomials up to a certain order. We fit the local phase gradient to the gradient of the sum of Zernike polynomials, which enables us to extract the polynomial coefficients and thus recover the phase of the beam at the atom plane. The fit gives us the correction that we have to apply on the SLM in order to have a flat wavefront at the atom plane. After applying the correction, we remeasure the wavefront with the same technique. In Fig. \ref{figM2}b-c, we show the Zernike contributions of the wavefront correction measurement for the above mentioned iteration and the expected improvements in the pointing of the rays after applying this correction on the SLM. \\

 The figure of merit is the spread of the pointing of the rays in the atom plane, which is shown in Fig. \ref{fig3}b-c before and after the wavefront optimisation. We iterate this procedure 17 times, as shown in Fig. \ref{fig3}d, slowly increasing the maximum degree of Zernike polynomial in the reconstruction, to correct for smaller wavefront deviations up to Zernike order 9. \\

We stop the procedure when we observe a convergence of the spread of the pointings in the atom plane. The results are limited by the spatial extent of the atomic cloud, which prevents us from determining precisely the correction for each patch (see Supplementary). \\

Tracking the ray directions at each iteration allows us to estimate the optimal wavefront correction, whose Zernike decomposition is shown in Fig. \ref{figM2}d. We attribute the high degree of coma and spherical aberration to an imperfect match of the optical axis of the cavity mirror and lenses during the optical contact procedure (see Supplementary).

\backmatter

\clearpage 

\bmhead{Acknowledgments}
We thank Michael Eichenberger for careful reading of the manuscripts. We acknowledge funding from
the Swiss National Science Foundation (Grant No. 184654), the Swiss State Secretariat for Education, Research and Innovation (Grants No. MB22.00063 and UeM019-5.1) and the Sandoz Family Foundation. FO and NR acknowledge support from QSIT INSPIRE Potentials Master Internship awards 2020 and 2021. \\
\newline

$\dag$ Current affiliation: Institute for Quantum Electronics, ETH Zürich, 8093 Zürich, Switzerland \\
$\ddag$ Current affiliation: Univ. Grenoble-Alpes, CEA, Leti, F-38000 Grenoble
\bibliographystyle{ieeetr}
\bibliography{microscope_paper_bibliography}

\clearpage 

\bmhead{Supplementary information}

\subsection{Optical Contact}\label{Suppl:Optical_Contact}

\begin{figure}[h]%
\centering
\includegraphics[width=0.5\textwidth]{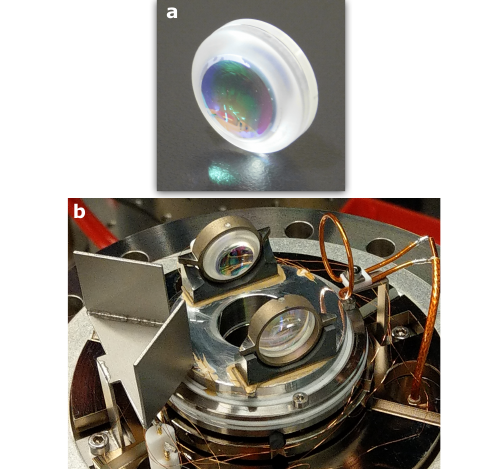}
\caption{\textbf{Pictures of the cavity-microscope.}  \textbf{a} Mirror-lens pair after optical contact. \textbf{b} Final cavity-microscope setup in vacuum: the contacted lens and mirrors are placed into round holders whose position is controlled by shear piezos. The cavity-microscope is mounted on a vibration damping platform \cite{Sauerwein_2022}. On the left of the picture, the atom shield prevents that the atoms from the oven coat the mirror surfaces, and on the right the radio frequency antenna is mounted \SI{2}{\cm} away from the cavity center. The setup is operated in ultra-high vacuum. }\label{figS2}
\end{figure}

Our cavity-microscope combines a high-finesse cavity with an on-axis microscope in a single optical device. This is accomplished by attaching two optical components, a mirror and a lens, together. The assembly process based on optical contact bonding starts by cleaning the flat surfaces of the mirror and the lens using the grad and wipe technique. Then, we mount them in an optical alignment setup to make sure that they were concentric with a pilot beam. To initiate the optical contact process, we placed a drop of water in the gap between the two optical elements, which attracted them together. After most of the water had evaporated, we removed the cavity lens assembly from the alignment setup and heated it to \SI{70}{\degreeCelsius} for \SI{1}{\hour} under vacuum.
Figure \ref{figS2} shows a picture of the cavity-microscope mounted in the vacuum chamber.

 \begin{figure}[h]%
\centering
\includegraphics[width=0.5\textwidth]{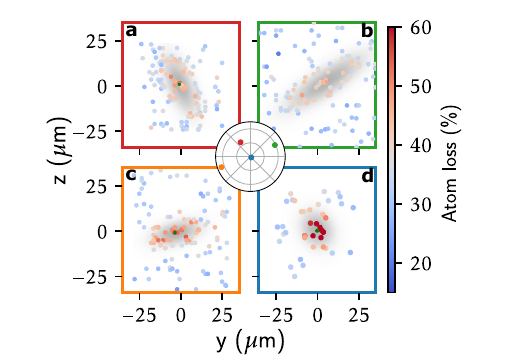}
\caption{\textbf{Example of ray tracing optimisation for the wavefront measurement.} \textbf{a-d}. For four different patches of the wavefront (central inset), we measure the atom losses when steering the ray on the (y,z) plane. We fit a 2D Gaussian profile to the data and we extract the phase gradient to apply on the SLM such that the ray hits the center of the atomic cloud. Due to the shape of the cloud, which is elongated in the cavity direction, the exact aspect ratio and orientation of the cloud depends on the position of the patch, as shown in \textbf{e} for the patch of panel b.}\label{figS3}
\end{figure}
\subsection{Optimisation algortihm}\label{secS4.9}
For the optimization of the wavefront of the control beam at the center of the cavity, we use a technique inspired from the Hartmann-Shack wavefront optimisation. In order to measure the wavefront  of the light on the atoms, we divide the area of the beam on the SLM in circular patches as described in Methods \ref{M:wavefront_engineering}. We steer the pointing of each patch in the atom plane by changing the phase gradient applied on the SLM. The goal is to find the optimal phase shift such that the ray hits the atomic cloud in the center. 
For each ray position, we measure atom losses after probing for \SI{50}{\micro\s} the cavity on resonance with the atomic transition and a cavity-probe detuning of $\Delta_{pc} =$ \SI{-10}{\mega\hertz}. \\

\begin{figure*}[h]%
\centering
\includegraphics[width=0.9\textwidth]{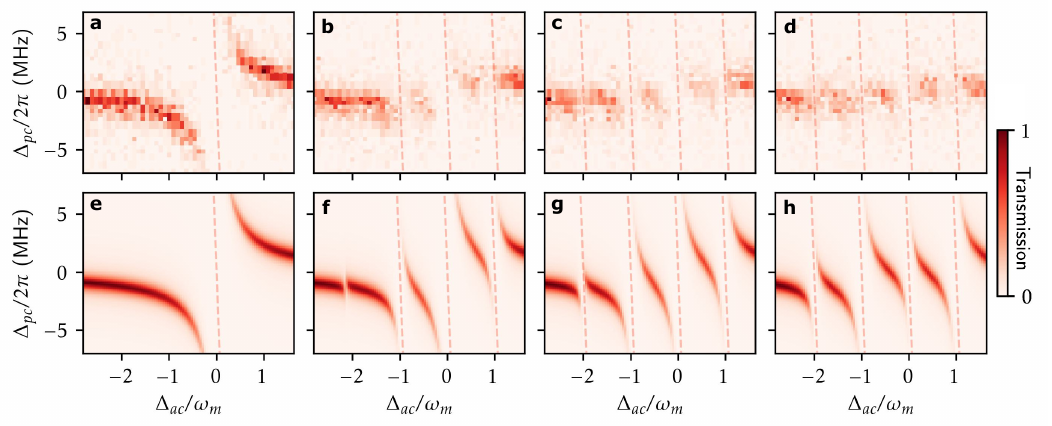}
\caption{\textbf{Direct simulation of the system homogeneously illuminated by the control beam.} Comparison between data (\textbf{a-d}) and direct simulation (\textbf{e-h}) for the data presented in Fig. \ref{fig2}e-h. From the simulation, we extract the vales of the maximum Rabi frequency of the control beam for each panel: from the left, $\Omega_0/2\omega_m = $ \qty{0}, \qty{0.9}, \qty{1.2}, \qty{1.8}.}\label{figS1}
\end{figure*}

We perform the search of the optimal pointing with a machine learning technique, using the MLOOP function of the Labscript Suite \cite{labscriptsuite2019}. Some examples of the optimization for different patches are shown in Fig. \ref{figS3}a-d.The search starts by sampling uniformly N points in this parameter space. After the points are acquired, we fit of a 2D Gaussian function to the data, knowing that the atomic cloud has an elongated shape in the cavity direction. The fit returns optimal parameters and their covariances. We choose the next sampling point according to where the uncertainty on our fit is maximal: we build an error matrix defined as $$E = J(\overrightarrow{x})\cdot C(\overrightarrow{x}) \cdot J(\overrightarrow{x})^{T}$$ where $J$ is the Jacoabian and $C$ is the covariance matrix of the fit and $\overrightarrow{x} = (y,z)$. We then extract N points to sample by drawing them from the error distribution. \\
We stop the optimisation once the fit error on the center of the 2D Gaussian is below \SI{0.4}{\micro\m}. The center of the Gaussian gives us the optimal phase gradient that we have to apply to the SLM patch such that the ray hits the atomic cloud in the center. Once all the patches are measured, we fit the wavefront as described in Methods \ref{M:wavefront_engineering}.

\end{document}